\documentclass[english,american,osajnl,twocolumn,showpacs,superscriptaddress,11pt,noamsmath]{revtex4-1}
\usepackage[T1]{fontenc}
\usepackage[latin9]{inputenc}
\usepackage{units}
\usepackage{textcomp}
\usepackage{bm}
\usepackage{amsmath}
\usepackage{graphicx}
\PassOptionsToPackage{version=3}{mhchem}
\usepackage{mhchem}
\usepackage{esint}

\makeatletter

    \@ifundefined{textcolor}{}
    {%
      \definecolor{BLACK}{gray}{0}
      \definecolor{WHITE}{gray}{1}
      \definecolor{RED}{rgb}{1,0,0}
      \definecolor{GREEN}{rgb}{0,1,0}
      \definecolor{BLUE}{rgb}{0,0,1}
      \definecolor{CYAN}{cmyk}{1,0,0,0}
      \definecolor{MAGENTA}{cmyk}{0,1,0,0}
      \definecolor{YELLOW}{cmyk}{0,0,1,0}
    }

\@ifundefined{showcaptionsetup}{}{%
 \PassOptionsToPackage{caption=false}{subfig}}
\usepackage{subfig}
\makeatother

\usepackage{babel}
\begin{document}

\title{Tomographic laser absorption spectroscopy \\
using Tikhonov regularization}

\author{Avishek Guha}

\affiliation{Merchant Gases R\&D, Air Products and Chemicals Inc., Allentown,
PA 18195, USA}

\author{Ingmar Schoegl}

\email{Corresponding author: ischoegl@lsu.edu}

\affiliation{Mechanical \& Industrial Engineering Department, Louisiana State
University, Baton Rouge, LA 70803, USA}
\begin{abstract}
The application of tunable diode laser absorption spectroscopy (TDLAS)
to flames with non-homogeneous temperature and concentration fields
is an area where only few studies exist. Experimental work explores
the performance of tomographic reconstructions of species concentration
and temperature profiles from wavelength-modulated TDLAS measurements
within the plume of an axisymmetric McKenna burner. Water vapor transitions
at 1391.67 nm and 1442.67 nm are probed using calibration free wavelength
modulation spectroscopy with second harmonic detection (WMS-\emph{2f}).
A single collimated laser beam is swept parallel to the burner surface,
where scans yield pairs of line-of-sight (LOS) data at multiple radial
locations. Radial profiles of absorption data are reconstructed using
Tikhonov regularized Abel inversion, which suppresses the amplification
of experimental noise that is typically observed for reconstructions
with high spatial resolution. Based on spectral datareconstructions,
temperatures and mole fractions are calculated point-by-point. Here,
a least-squares approach addresses difficulties due to modulation
depths that cannot be universally optimized due to a non-uniform domain.
Experimental results show successful reconstructions of temperature
and mole fraction profiles based on two-transition, non-optimally
modulated WMS-\emph{2f} and Tikhonov regularized Abel inversion, and
thus validate the technique as a viable diagnostic tool for flame
measurements.
\end{abstract}


\maketitle

\section{Introduction}

Over the last couple of decades, a significant amount of work has
been performed in the area of temperature and species concentration
determination using tunable diode laser absorption spectroscopy \citep{Philippe1993,Zhou2003,Zhou2005,Farooq2008a}.
Line-of-sight TDLAS signals are, in essence, attenuation that is integrated
along the beam path. Hence, these techniques, -- while good for domains
with uniform temperature and species concentration, -- are not readily
applicable for areas where the fields vary significantly. For such
cases, the application of tomography along with TDLAS provides a means
to achieve spatially resolved beam attenuation data which can be subsequently
used for the determination of temperature and concentration values. 

Tomographic TDLAS, for 1D tomography in particular, is generally achieved
by acquiring LOS attenuation data, -- i.e. projections, -- using a
parallel rake of rays scanned across an axisymmetric flame \citep{silver1995,Villarreal2005}.
Projections are subsequently deconvolved using a numerical implementation
of Abel's inversion equation \citep{Abel1826,Dasch1992,Villarreal2005}.
Since Abel inversion is an inherently ill-conditioned problem \citep{Dasch1992,Daun2006},
small levels of measurement noise in projection data are amplified
in the solution. Therefore, deconvolutions are usually performed for
coarse grid spacing \citep{Dasch1992} and/or after artificial smoothing
of projection data \citep{silver1995,Villarreal2005}.

Available work on tomographic TDLAS falls into two categories: tomography
of axisymmetric domains, -- which is essentially 1D, -- and reconstructions
of 2D slices using projections from multiple directions/views. Another
distinction can be made with respect to the number of spectral transitions
being probed. Two important prior studies on 1D tomographic TDLAS
are works by Silver et al. \citep{silver1995} and Villarreal and
Varghese \citep{Villarreal2005}. Both studies use traditional Abel
inversion to deconvolve LOS data and use relatively large beam spacing.
While the experimental setup of the present study is a variation of
the one used by Villarreal and Varghese \citep{Villarreal2005}, the
previous study used 9 spectral transitions to simultaneously reconstruct
temperature and concentration profiles using a scanned direct absorption
method. Silver et al. \citep{silver1995} used WMS-\emph{2f} to reconstruct
concentration profiles of $\mathrm{H_{2}O}$ for a flame in microgravity.
While this approach is nominally similar, the previous study decoupled
measurements by a semi-computational approach: concentrations are
found directly from lines with low temperature sensitivity, whereas
corresponding temperatures are calculated. The present work, however,
determines concentration (mole fractions) and temperature simultaneously
and furthermore seeks to augment the performance of highly resolved
Abel inversion by the introduction of Tikhonov regularization. An
alternative approach is hyperspectral tomography, which was introduced
in the context of 2D reconstructions \citep{ma2008,ma2008b,cai2008,Ma2009,hagen2007}.
In this method, LOS absorbance data are obtained for a large number
of spectral transitions using a broadly tunable diode laser. The strength
of this method is that the redundancy of spectral information allows
for a reduction of views for a satisfactory reconstruction. Furthermore,
recent proof-of-concept computational work indicates that hyper-spectral
LAS methods can be combined with WMS-2f \citep{cai2014b}. Hyper-spectral
methods require frequency-agile lasers that allow for broad-band modulation.
In 1D tomography, the redundancy of spectral information is less critical;
as will be shown, a small number of inexpensive laser diodes with
narrow modulation capability is sufficient.

The goal of this study is to produce high resolution 1D reconstructions
using Tikhonov regularized Abel inversion with two-transition WMS-\emph{2f}
TDLAS, where experimental validations are performed for an axisymmetric
McKenna burner. WMS-\emph{2f} for species measurements dates back
to at least the early 1990's \citep{silver1992,Philippe1993,silver1995}
and has found application in a range of recent studies \citep{Liu2004b,Liu2005,Li2006,Farooq2009b}.
Compared to direct absorption (DA) spectroscopy, WMS-\emph{2f} has
the advantage of producing a significantly higher signal-to-noise
ratio. This is achieved by detecting the harmonics at frequencies
that are high enough to reject much of the low-frequency high-amplitude
relative intensity noise \citep{Kluczynski1999,Farooq2009b} that
are typical in diode lasers. From an experimental point of view, WMS-\emph{2f}
is advantageous \citep{Liu2006,Rieker2009} since the first harmonic
signal can be used as an effective tool for normalizing any variation
of the second harmonic signal (calibration-free) \citep{Rieker2009,Li2006}
for optically thin samples. This approach is attractive, as it eliminates
low-frequency noise and makes measurement independent of signal disturbances
due to window fouling or beam walking. Since temperature and concentration
vary significantly within the domain, modulation depths for WMS-\emph{2f}
cannot be universally optimized as is commonly done for homogeneous
media \citep{Liu2004b,Farooq2009b}. Instead of a two-transition ratio
technique with optimal modulation depths, temperature and mole fractions
are simultaneously determined by using a Levenberg-Marquardt least
squares algorithm with non-optimal modulation depths.

\section{Theory}

A wavelength-modulated laser beam passing through a non-uniform temperature
and concentration field is subject to path-integrated beam attenuation.
The resulting signal can be deconvolved using traditional tomographic
methods to produce local contributions to the beam attenuation. These
values are then used to calculate temperature and mole fraction via
a two-transition WMS-\emph{2f} method. The following sections summarize
the theory behind WMS-\emph{2f} thermometry and detail Tikhonov regularized
Abel inversion of data derived from WMS-\emph{2f}.

\subsection{Wavelength Modulation Spectroscopy}

For a monochromatic beam passing through an absorbing medium, Beer-Lambert's
law relates the incoming (\foreignlanguage{english}{$I_{0}$}) and
transmitted ($I_{t}$) intensity at wavelength $\nu$ as 
\begin{equation}
\left(\frac{I_{t}}{I_{0}}\right)_{\nu}=\exp(-\alpha(\nu))=\tau(\nu)\label{eq:Beer-Lambert}
\end{equation}
Here, $\tau$ is the transmission coefficient while $\alpha$ is the
absorbance. For uniform temperature and concentration, the absorbtion
due to a single species is 
\begin{equation}
\alpha(\nu)=PXL{\displaystyle \sum_{i}}S_{i}(T)\phi_{i}(\nu,X)
\end{equation}
where $P$ is the pressure, $X$ is the mole fraction of the absorbing
species, $L$ is the total path length, $S_{i}$ is the temperature
dependent line-strength of the $i$-th transition, and $\phi_{i}$
is the corresponding Voigt lineshape function. 

In WMS-\emph{2f} experiments, a targeted absorption transition is
scanned using a tunable diode laser, generally by a low frequency
sawtooth ramp ($f_{s}\approx1\,\mathrm{kHz}$) superimposed with a
high frequency modulation ($f\approx150-200\,\mathrm{kHz}$). The
resulting instantaneous laser frequency is given by
\begin{equation}
\nu(t)=\bar{\mathrm{\nu}}_{s}(t)+a\,\cos(\omega t)\label{eq:freq eqn-1}
\end{equation}
Here, $\bar{\nu}_{\mathrm{s}}\mathrm{(t)}$ is the slowly ramping
central laser wavelength, whereas $a$ and $\omega=2\pi f$ are modulation
amplitude and angular frequency of laser frequency modulation (FM),
respectively. In the following, $\mathrm{\theta=\omega t}$ is introduced
for convenience, and the slowly varying $\bar{\mathrm{\nu}}_{s}(t)\approx\bar{\nu}$
is considered constant with respect to the rapid modulation due to
$f_{s}\ll f$.

Because of the modulation in wavelength, the instantaneous value of
the transmission coefficient can be expressed as a periodic even function
(or a cosine series) in $\theta$, which takes the form
\begin{equation}
\tau(\nu(t))=\tau\left(\bar{\nu}+a\,\cos\theta\right)=\sum_{k=0}^{\text{\ensuremath{\infty}}}H_{k}\mbox{\,}\cos k\theta\label{eq:cosine-series}
\end{equation}
Here, the coefficients $H_{k}$ for the $k$-th harmonic ($\mathrm{k=0,1,2},\dots$)
are found as 
\begin{equation}
H_{k}=\frac{1}{n\pi}\int_{-\pi}^{+\pi}\tau(\bar{\nu}+a\,\cos\theta)\cos k\theta d\theta\label{eq:Hk_hom}
\end{equation}
where $n=2$ if $k$ is zero and $n=1$ otherwise. 

Under the assumption of an optically thin transition (i.e. $\alpha(\nu)<0.1$
\citep{Li2006}), Eq. \ref{eq:Beer-Lambert} is linearized as $\tau(\nu)\text{ \ensuremath{\approx}}1-\alpha(\nu)$.
Substituting into Eq. \ref{eq:Hk_hom} yields\begin{subequations}

\begin{equation}
H_{k}=PXL\sum_{i}S_{i}(T)\Phi_{ik}(X)\label{eq:Hk_inhom}
\end{equation}
with line-shape integrals $\Phi_{ik}$ defined as 
\begin{equation}
\Phi_{ik}(X)=-{\textstyle \frac{1}{n\pi}\int_{-\pi}^{+\pi}\phi_{i}(\nu(\bar{\nu},a,\theta),X)\cos k\theta d\theta}\label{eq:lineshape_integral}
\end{equation}
\end{subequations}In general, $\Phi_{ik}$ are relatively insensitive
to composition if an appropriate modulation depth $a$ is chosen.
For moderate ($\mathrm{\pm10\%}$) variations of concentration, a
modulation amplitude $a$ close to 2.2 half-widths of the transition
yields almost constant $\Phi_{ik}$ \citep{Liu2004b,Farooq2009b}.
However, this property cannot be used in the context of tomography.

For the general case where gas composition and temperature vary along
the line of sight $L$, the path-integrated harmonic coefficients
are rewritten as\begin{subequations}\label{eq:LOSharmonics}
\begin{eqnarray}
H_{k} & = & \int_{0}^{L}\! h_{k}(T(\ell),X(\ell))\, d\ell\label{eq:integratedLOS}
\end{eqnarray}
where the local contribution to LOS values are
\begin{equation}
h_{k}(T,X)=P\, X\;{\textstyle \sum_{i}S_{i}(T)\Phi_{ik}(X)}.\label{eq:localLOS}
\end{equation}
\end{subequations}Equation \ref{eq:LOSharmonics} illustrates that
individual harmonics in WMS-\emph{2f} are path-integrated quantities,
and thus can be deconvolved using conventional tomography. It is,
however, noted that the assumption of an optically thin medium is
essential to this approach.

\begin{figure}
\centering{}\includegraphics[width=0.7\columnwidth]{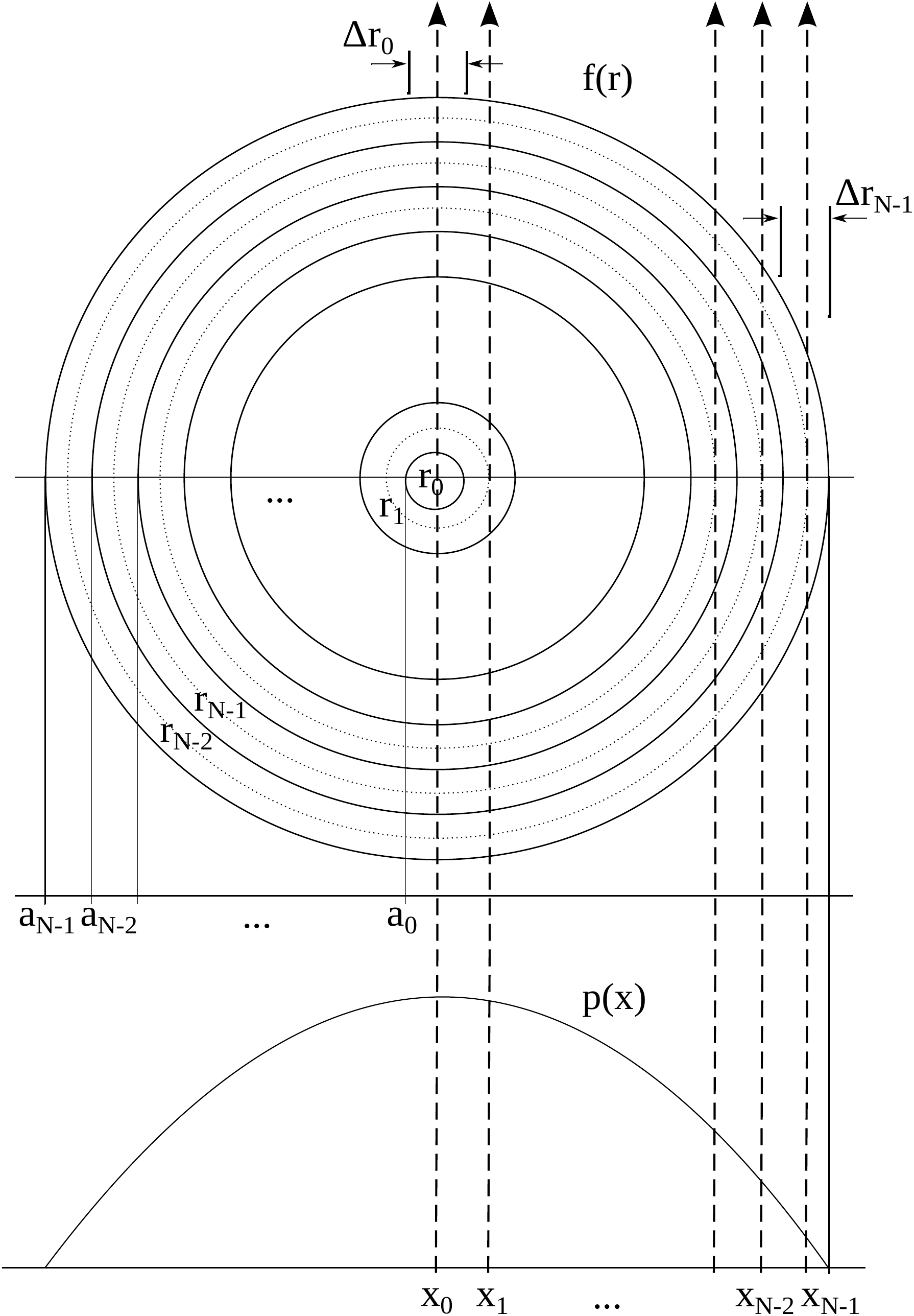}\protect\caption{Domain discretization for Abel inversion. \label{fig:Domain}}
\end{figure}

\subsection{Tomography of Spectroscopic Data\label{sub:Tomography-theory}}

The deconvolution of path-integrated WMS-\emph{2f} signals produces
spectral data that are a function of local temperature and concentration.
Here, Tikhonov regularized Abel inversion \citep{Daun2006} addresses
the inherent ill-conditioned nature of Abel inversion with closely
packed rays. In the following, the concept is summarized for an arbitrary
radially changing field variable $f(r)$ that result in projections
$p(x)$. In the case of WMS-\emph{2f}, these values correspond to
linear combinations of local harmonics \foreignlanguage{english}{$h_{k}(T,X)$}
and path-integrated measurements \foreignlanguage{english}{$H_{k}$}
at radial locations $r$, respectively. A detailed discussion of measured
signals is given in Section \ref{sub:Signal-Processing}.

For a numerical implementation of Abel inversion, an axisymmetric
domain is subdivided into annular rings with rays passing through
the center of each annulus as illustrated in Figure \ref{fig:Domain}.
Thus, rays $x_{0},\, x_{1},...,x_{N-1}$ pass through annuli with
labels $a_{0},\, a_{1},...,a_{N-1}$, respectively, where $x_{i}$
correspond to radial distances $r_{i}$. With this definition of rays
and annuli in place, the projection of field variables can be written
as a sum of integrals over the whole domain as \begin{subequations}
\begin{equation}
\begin{aligned}p(x_{m})= & \,2\,\mathrm{\sum}_{n=m}^{N-1}\int_{b_{n}}^{a_{n}}\frac{f(\tilde{r})\tilde{r}}{\left(\tilde{r}^{2}-r_{m}^{2}\right)^{\nicefrac{1}{2}}}d\tilde{r}\end{aligned}
\label{eq: Discr proj}
\end{equation}
where the lower integration limit is
\[
b_{n}=\left\{ \begin{array}{cc}
r_{n}, & \ n=m\\
a_{n-1}, & \ n>m
\end{array}\right.
\]
If the field variable $f(r)$ is approximated by a Taylor series expansion
around $r_{i}$, all integrals can be precalculated, where the Abel
3-point (ATP) scheme is used \citep{Dasch1992,Villarreal2005}. Thus,
the system of analytical equations (Eq. \ref{eq: Discr proj}) is
represented by
\begin{equation}
\mathbf{\mathbf{A}}\mathbf{f}=\mathbf{p}\label{eq:proj eq}
\end{equation}
\end{subequations}where the vectors $\mathbf{f}=[f_{0},f_{1},\dots f_{N-1}]^{T}$
and $\mathbf{p}=[p_{0},p_{1},\dots p_{N-1}]^{T}$ contain discretized
field variable values and projection values, respectively. The matrix
$\mathrm{\mathbf{A}}$ acts as the \emph{projection matrix}, where
elements $A_{mn}$ define the contribution of the $n$-th annulus
to the projection value of the $m$-th ray. 

Tikhonov regularization addresses the inherent ill-conditioning of
the projection matrix $\mathrm{\mathbf{A}}$ \citep{Hansen1987,Daun2006},
which causes the amplification of measurements noise in the solution
\citep{Dasch1992,Daun2006}. Destabilization of reconstructed solution
in highly resolved grids can be mitigated by augmenting the information
by an additional set of equations 
\begin{equation}
\lambda\mathbf{L}\mathbf{f}=\bm{0}\label{eq:reg}
\end{equation}
Here, $\lambda$ is a parameter that controls the extent of regularization,
whereas $\mathbf{L}$ is used to penalize large gradients in the solution
and is implemented as a discrete version of the $\nabla$ operator,
i.e.
\[
\mathbf{L}=\left[\begin{array}{ccccc}
1 & -1 & 0 & . & 0\\
0 & 1 & -1 & . & 0\\
0 & 0 & . & . & .\\
. & . & . & . & -1\\
0 & 0 & . & 0 & 1
\end{array}\right]
\]
Equations \ref{eq:proj eq} and \ref{eq:reg} form an overdetermined
system, where a least-squares solution is sought as
\begin{equation}
\bm{\mathbf{\mathrm{f}}}_{\lambda}=\arg\,\min\left\{ \left\Vert \left[\begin{array}{c}
\mathbf{A}\\
\lambda\mathbf{L}
\end{array}\right]\mathbf{f}-\left[\begin{array}{c}
\mathbf{p}\\
\bm{0}
\end{array}\right]\right\Vert \right\} \label{eq:tikhonov}
\end{equation}
Here, the regularization parameter $\lambda$ controls the relative
weight placed on accuracy and smoothness of the solution: a small
$\lambda$ value implies a solution that traces measurement points
(including noise) accurately but may be highly oscillatory, whereas
a large $\lambda$ enforces a smooth solution that may deviate from
measurements. While multiple methods for a proper choice of $\lambda$
exist, this work adopts the \emph{L-curve} criterion \citep{Hansen1987},
which yields $\lambda\approx\mathcal{O}(1)$, and $\lambda=1$ is
used for all reconstructions.

\begin{figure}
\centering{}\includegraphics[width=1\columnwidth]{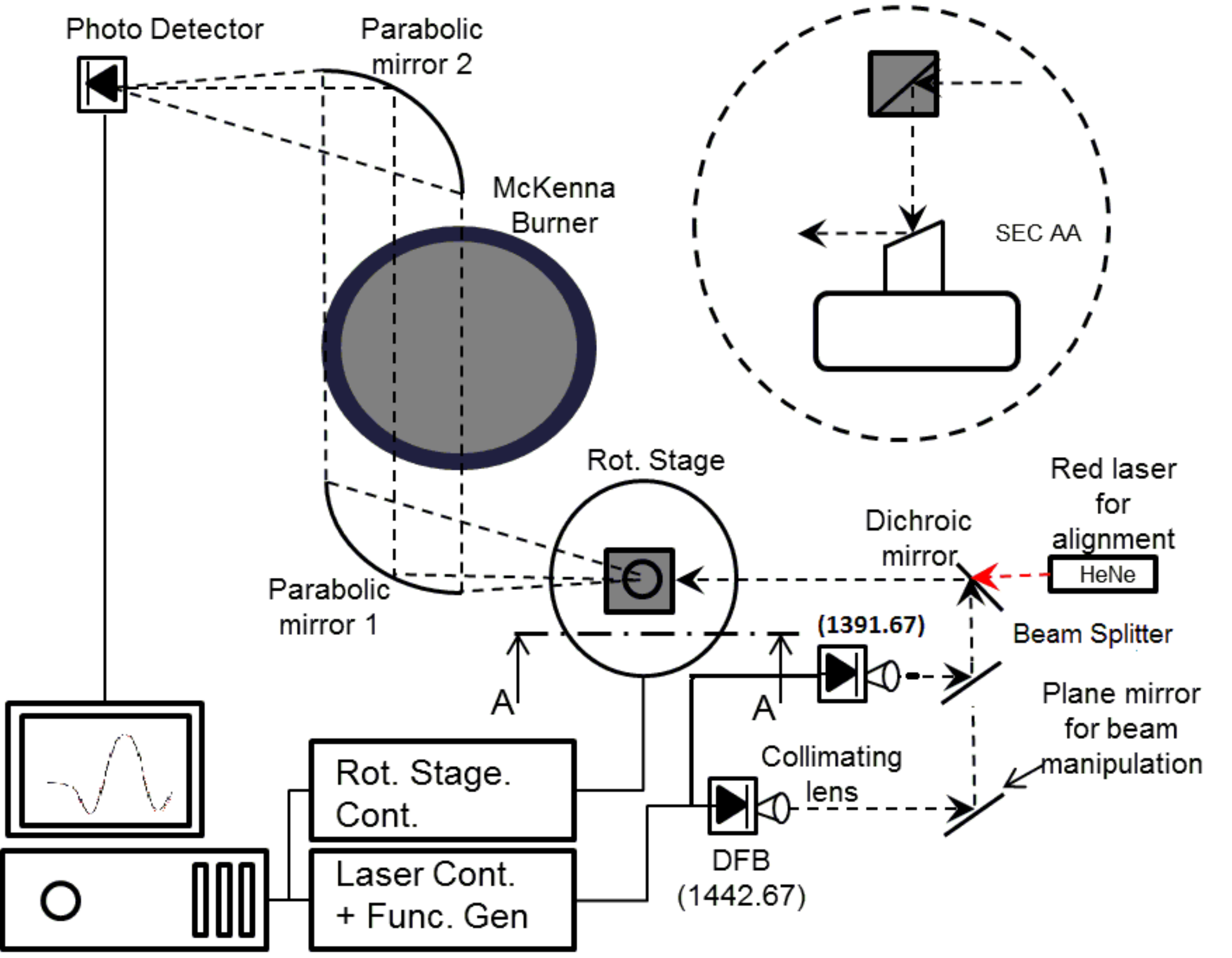}\protect\caption{Experimental setup. \label{fig:Setup}}
\end{figure}

\section{Experimental Methods}

\subsection{Experimental Setup}

Figure \ref{fig:Setup} shows the experimental setup that is used
for 1D tomographic TDLAS of the post-flame zone of an axisymmetric
McKenna burner (Holthuis \& Associates). Spectroscopic data is generated
by a pair of multiplexed laser beams tuned to transitions at 1391.67
nm and 1442.67 nm, respectively. The test section is probed by parallel
rays, which are swept parallel to the burner surface with a setup
consisting of a periscope mounted on a rotation stage and a single
detector, both of which are positioned in the focal points of a pair
of parabolic mirrors \citep{Villarreal2005}. Injection currents and
temperatures of both diode lasers (NLK1E5C1TA, NTT Electronics) are
controlled by a dedicated controller (LDC-3908, ILX Lightwave). The
temperature controller is used to set the central laser emission wavelength
while the injection current is modulated using a programmed function
generator (CG4340, Gagecard). The TO can diode lasers are collimated
using an aspheric lens and shaped using an iris before being multiplexed
in the time domain. The beam diameter is 0.5 mm, which also corresponds
to the spatial resolution. The movement of a high-precision rotation
stage (RGV-100 BL, Newport) is controlled by LabVIEW (National Instruments)
to produce measurements in increments of 1mm. The raw signal from
the photodetector (PDA-10CS, Thorlabs) is digitized using a digital
oscilloscope (PCI-5105, National Instruments). 

The function generator uses a 200Hz scanning ramp super-imposed with
a 50 kHz sinusoidal modulation to create harmonics of the absorption
signal. In measurements, sets of 10 sawtooth ramps (time-multiplexed,
i.e. five per transition) are captured and passed individually through
a digital lock-in-amplifier (implemented in software/LabVIEW). Harmonics
of the modulated signal (\emph{1f} and \emph{2f}) are written to a
binary file for further post-processing. At each radial location,
100 data sets are obtained. Due to oversampling, a measurement at
an individual location takes 2 seconds per line. Modulation depths
are chosen as $0.13\,\mathrm{cm^{-1}}$ and $0.1\,\mathrm{cm^{-1}}$
for transitions at 1442.67 nm and 1391.67 nm, respectively. The modulated
laser output is validated using a Silicon etalon (Lightmachinery Inc.)
with a free spectral range (FSR) of $\mathrm{\sim0.017\, cm^{-1}}$
in the $1.4\,\mu\mathrm{m}$ range. 

Experiments use a methane/air McKenna flame with a nominal radius
of $30\,\mathrm{mm}$ and an external shroud thickness of $\mathrm{7.5\, mm}$.
The burner produces a stationary flat flame, where equivalence ratios
of 0.7 and 0.8 are tested for a mixture flow velocity of 20 cm/s.
Tomographic measurements are taken 7mm above the burner plate. Flow
rates of methane and air are set by mass flow controllers (32907-71/32907-73,
Cole-Parmer) using a LabVIEW interface. The flame is shielded by a
nitrogen shroud flow, where the velocity is matched with the cold
air-fuel mixture in order to minimize shear at the edge of the flame.
The McKenna burner is supplied with cooling water at a rate of 800
ml/min, where the inlet temperature is held at $\mathrm{4\,\text{\textdegree}C}$
by an enclosed ice bath.

\subsection{Signal Processing\label{sub:Signal-Processing}}

The non-ideal intensity modulation (IM) due to injection current modulation
of a typical diode laser is modeled as \citep{Reid1981,Li2006} 
\begin{equation}
I_{0}(\theta)=\bar{I}_{0}\left(1+i_{0}\cos(\theta+\psi_{1})+i_{2}\cos(2\theta+\psi_{2})\right)\label{eq:Laser output}
\end{equation}
Here, $\bar{I}_{0}$ is the average incident laser intensity at $\bar{\nu}$,
whereas $i_{\mathrm{0}}$ and $i_{2}$ are normalized linear (1f)
and non-linear (2f) modulation amplitudes, respectively. The corresponding
phase shifts are given by $\psi_{1}$ and $\psi_{2}$. Both are FM/IM
phase shifts which gauge the lag between wavelength and intensity
modulation.

In the following, the gas-attenuated raw detector signal is expressed
in terms of instantaneous incident laser intensity and transmission
coefficient as
\begin{equation}
I_{t}(\theta,\bar{\nu},a)=GI_{0}\left(\theta\right)\ \tau(\bar{\nu}+a\cos\theta)\label{eq:detector signal}
\end{equation}
 where G is the optical/electrical gain of the system. 

In order to find the $k$-th overtone of the signal, the raw detector
signal is passed through a digital lock-in-amplifier, where the detector
signal is multiplied by reference signals (either $\cos(k\theta+\psi_{d})$
or $\sin(k\theta+\psi_{d})$) to produce $x$ and $y$-components.
Here, the detection phase shift $\psi_{d}$ is the shift between intensity
modulation and reference signal.

Traditionally, background subtracted lock-in-amplifier outputs are
assessed as the vector sum of $x$ and $y$ components \citep{Li2006}.
While this approach is independent of $\psi_{d}$, the resulting values
are not linear combinations of harmonics and therefore cannot be used
for tomography. In the present work, this issue is resolved by adjusting
$\mathrm{\psi_{d}}$ to zero, which allows for traditional tomography
based on the $x$-component of the signal alone. In the following,
1f and 2f outputs produced by the lock-in-amplifier are denoted by
$S_{\mathit{1f}}$ and $S_{\mathit{2f}}$, respectively.

It can be shown that both $S_{\mathit{1f}}$ and $S_{\mathit{2f}}$
are directly proportional to the instantaneous mean intensity of the
laser \citep{Li2006}. The 1f signal is given by 
\begin{equation}
S_{\mathit{1f}}={\textstyle \frac{1}{2}}G\bar{I}_{0}i_{0}\left(1-f(P,X,L,T)\right)\approx{\textstyle \frac{1}{2}}G\bar{I}_{0}i_{0}\label{eq:1f_signal}
\end{equation}
where $S_{1f}$ scales with laser intensity and 1f modulation amplitude
under the assumption of optically thin samples where $f(P,X,L,T)\ll1$.
The calibration-free WMS-\emph{2f} method uses this property to normalize
the 2f signal by the 1f signal as
\begin{eqnarray}
S_{\mathit{2f}}/S_{\mathit{1f}} & = & \frac{1}{i_{0}}[H_{2}+{\textstyle \frac{1}{2}}i_{0}(H_{1}+H_{3})\cos\psi_{1}+\dots\nonumber \\
 &  & +\, i_{2}\left(1+H_{0}+{\textstyle \frac{1}{2}}H_{4}\right)\cos\psi_{2}]\label{eq:1fnorm2f}
\end{eqnarray}
where $H_{k}$ are defined in Eq. \ref{eq:integratedLOS}. 

In the absence of absorption, all $H_{k}$ are zero, and the measured
signal represents the background signal due to non-linearity of the
diode laser, i.e. 
\begin{equation}
S_{\mathit{2f},0}/S_{\mathit{1f}}=\frac{i_{2}}{i_{0}}\cos\psi_{2}\label{eq:RAM}
\end{equation}
which is also known as residual amplitude modulation or RAM \citep{Li2006,Kluczynski1999,Kluczynski2001}.
As $H_{0},H_{4}\ll1$ in the optically thin limit, RAM is the leading
order term among non-linear (\emph{2f}) modulation effects. While
RAM is usually neglected for small amplitudes of modulation \citep{Li2006},
it has to be considered if signal strengths are of similar orders
of magnitude.

\begin{figure}
\centering{}\includegraphics[width=0.9\columnwidth]{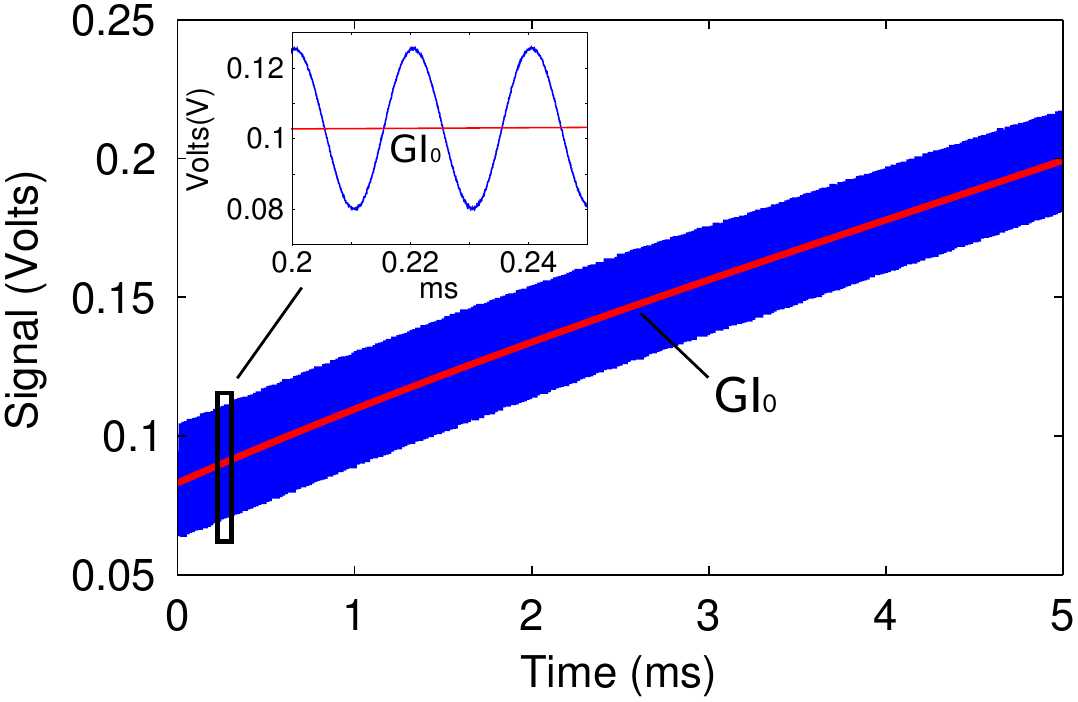}\protect\caption{Raw detector signal used for laser characterization: raw signal for
determination of $i_{0}$.\label{fig:laser_char}}
\end{figure}

\subsection{Laser Characterization}

In Equations \ref{eq:1fnorm2f} and \ref{eq:RAM}, linear $(i_{0},\,\psi_{1})$
and non-linear $(i_{2},\,\psi_{2})$ modulation parameters are specific
to the laser. Apart from spectroscopic data, measurements do not require
further calibration once a laser diode is characterized.

In scanned WMS-\emph{2f}, the injection current results from a high-frequency
modulation that is super-imposed on a low frequency ramp that determines
the slowly varying center wavelength $\bar{\nu}$. In order to find
$i_{0}$, this output was normalized by the central detector output,
which is found by fitting a $\mathrm{2^{nd}}$ order polynomial as
shown in Fig. \ref{fig:laser_char}. The normalized linear amplitude
$i_{0}$ is then found by performing a best-fit of a sinusoidal curve
with the same frequency to the intensity signal and normalizing it
with the average signal strength \citep{Li2006}. The linear FM/IM
phase shift $\psi_{1}$ is determined by simultaneous monitoring of
the output of two beam arms, where one is measured directly and the
other is passed through an etalon. 

Non-linear modulation effects due to $i_{2}$ and $\psi_{2}$ result
in RAM (Eq. \ref{eq:RAM}), which becomes non-negligible near the
edge of the flame. In this work, RAM is considered, while higher-order
non-linear modulation terms are neglected. In experiments, LOS signals
are recorded without the presence of a flame and subtracted from measurements.
In absence of absorption, this signal is equivalent to RAM. Absorption
by ambient air is, however, non-negligible at 1391.67nm, where corrections
account for both outside path length and RAM.

\subsection{Tomographic TDLAS}

In tomographic TDLAS, the output of the lock-in-amplifier is recorded
for two spectral transitions at multiple radial locations. Resulting
spectroscopic data, -- i.e. $(S_{\mathit{2f}}-S_{\mathit{2f},0})/S_{\mathit{1f}}$
representing linear combinations of $H_{k}$, -- are then deconvolved
according to Equation \ref{eq:tikhonov}. Thus, the deconvolved spectral
data are linear combinations of $h_{k}(T,X)$ (Eq. \ref{eq:localLOS}).
In the following, deconvolved measurement data are referred to as
$\zeta_{j}(T,X)$, where $j=1,2$ correspond to the two probed transitions.

In the context of tomography, variations of concentrations are not
moderate, which means that line-shape-integrals (Eq. \ref{eq:lineshape_integral})
cannot be approximated as constants as commonly assumed in WMS-\emph{2f}
\citep{Liu2004b,Farooq2009b}. Instead, a Levenberg-Marquardt scheme
with temperature and mole fraction as free parameters is adopted to
account for non-ideal behavior. Here, a variable $\mathrm{\Delta}$
is defined as 
\begin{equation}
\Delta(T,X)=\sqrt{{\textstyle \sum_{j=1,2}\left(\bar{\zeta}_{j}(T,X)/\zeta_{j}-1\right)^{2}}}\label{eq:LM}
\end{equation}
where $\bar{\zeta}_{j}(T,X)$ represent precalculated data for simulated
measurements. Solutions for $T$ and $X$ are found by minimizing
$\Delta(T,X)$ at each radial location using gradient-based optimization.

Simulated measurements are stored in a look-up-table, which contains
spectral absorbance calculations for spectral data and gas composition
of interest. Spectral parameters are based on HITRAN and HITEMP databases
\citep{rothman2009,rothman2010}, where transitions with line-strengths
less than $10^{-34}\,\mathrm{cm^{-1}/(molecule\, cm^{-2})}$ are neglected.
Simulated measurement data for \foreignlanguage{english}{$\bar{\zeta}_{j}(T,X)$}
are obtained by taking a fast-Fourier transform (FFT) of the absorbance
line shape around the transition line-centers where proper experimental
values for $a$, $i_{0}$ and $\psi_{1}$ are substituted.

\paragraph*{Uncertainties.}

Reconstructions are created based on 100 data sets per location, where
each set contains 5 sawtooth scans per transition. For each sawtooth
scan, WMS-\emph{2f} signals are processed and mean values and standard
variations are recorded. Uncertainties based on the standard variation
($\pm1\sigma$) are propagated to uncertainties in $\zeta_{i}(T,X)$
via the Abel uncertainty equation \citep{Dasch1992}. The projection
matrix for Tikhonov regularized Abel inversion is
\begin{equation}
A^{\#}=(A^{T}A+\lambda^{2}L^{T}L)^{-1}A^{T}
\end{equation}
Uncertainty in $\zeta_{j}(T,X)$ are subsequently used to calculate
the uncertainties in the temperature-mole fraction surface described
by Eq. \ref{eq:LM}. Additional uncertainties due to mass-flow controllers
($\pm2\%$) do not affect measurements directly, and thus are only
of concern for the validation process.

\section{Results and Discussions}

\begin{figure}
\begin{centering}
\subfloat[Projection data at 1442.67 nm.\label{fig:proj_line1}]{\centering{}\includegraphics[width=0.9\columnwidth]{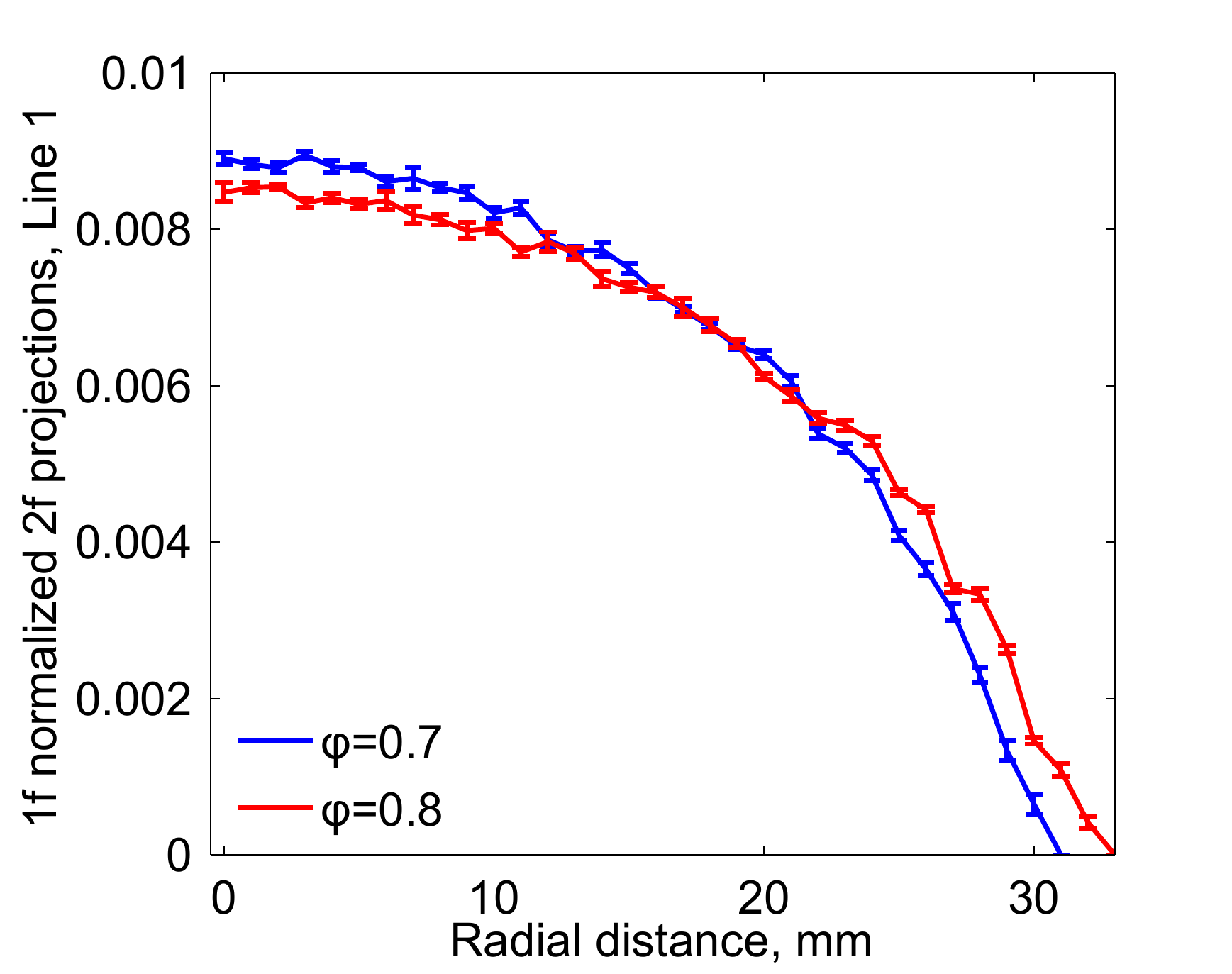}}
\par\end{centering}

\centering{}\subfloat[Projection data at 1391.67 nm.\label{fig:proj_line2}]{\centering{}\includegraphics[width=0.9\columnwidth]{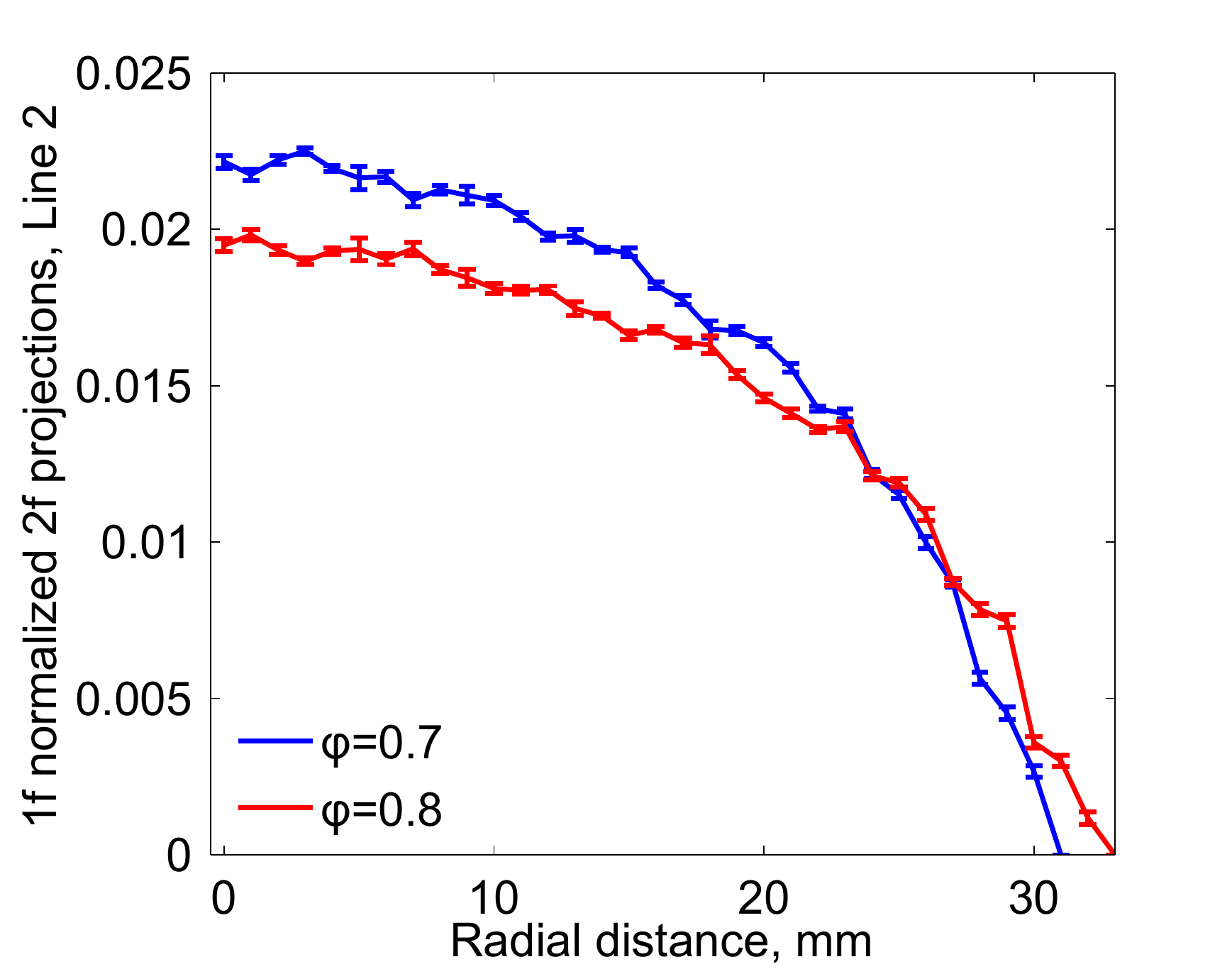}}\protect\caption{Projection data for spectral line measurements at $\phi=0.7$ (blue)
and $0.8$ (red). \label{fig:Result_proj}}
\end{figure}

\begin{figure}
\begin{centering}
\subfloat[Reconstructed profile at 1442.67 nm.\label{fig:line1}]{\centering{}\includegraphics[width=0.9\columnwidth]{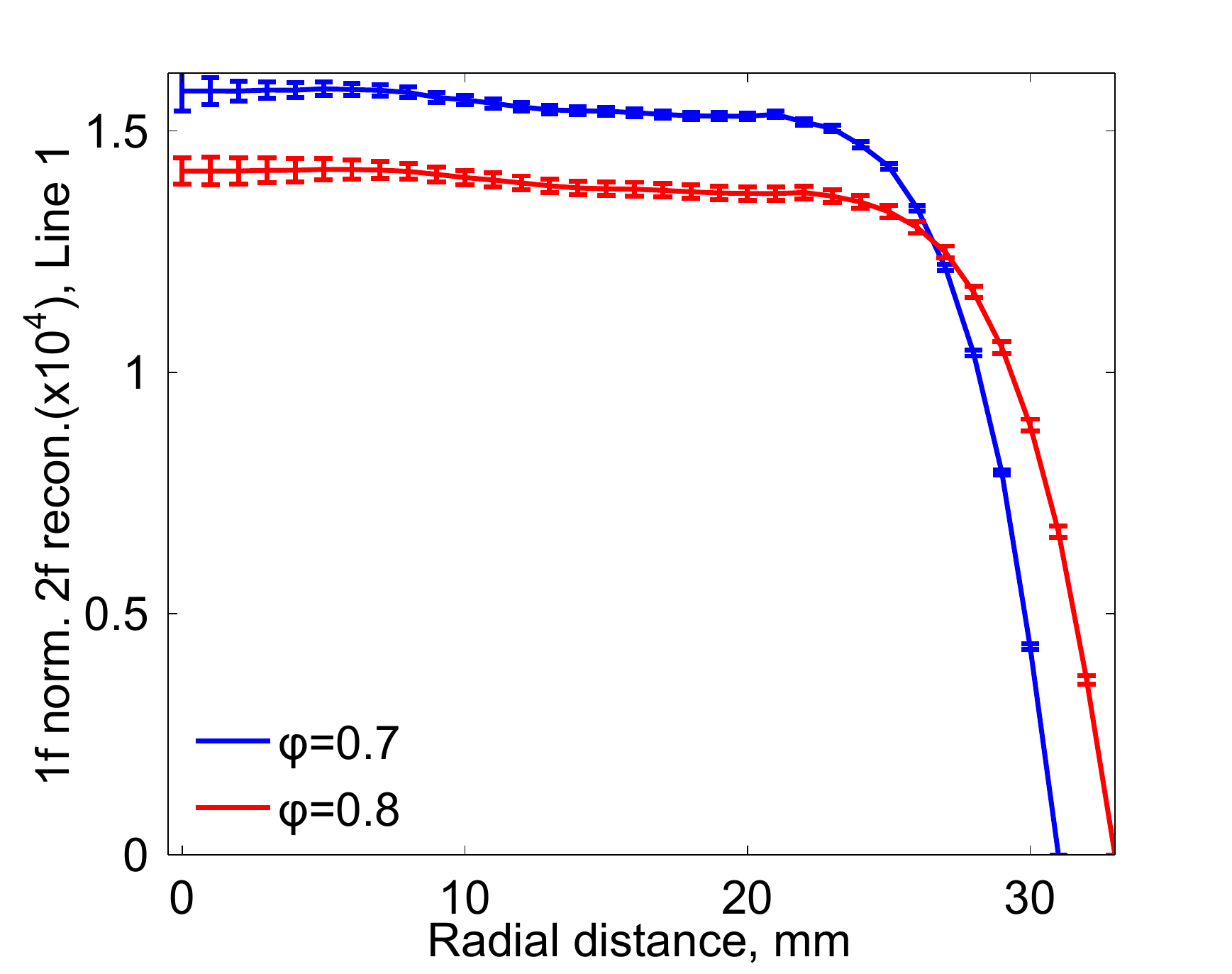}}
\par\end{centering}

\centering{}\subfloat[Reconstructed profile at 1391.67 nm.\label{fig:line2}]{\centering{}\includegraphics[width=0.9\columnwidth]{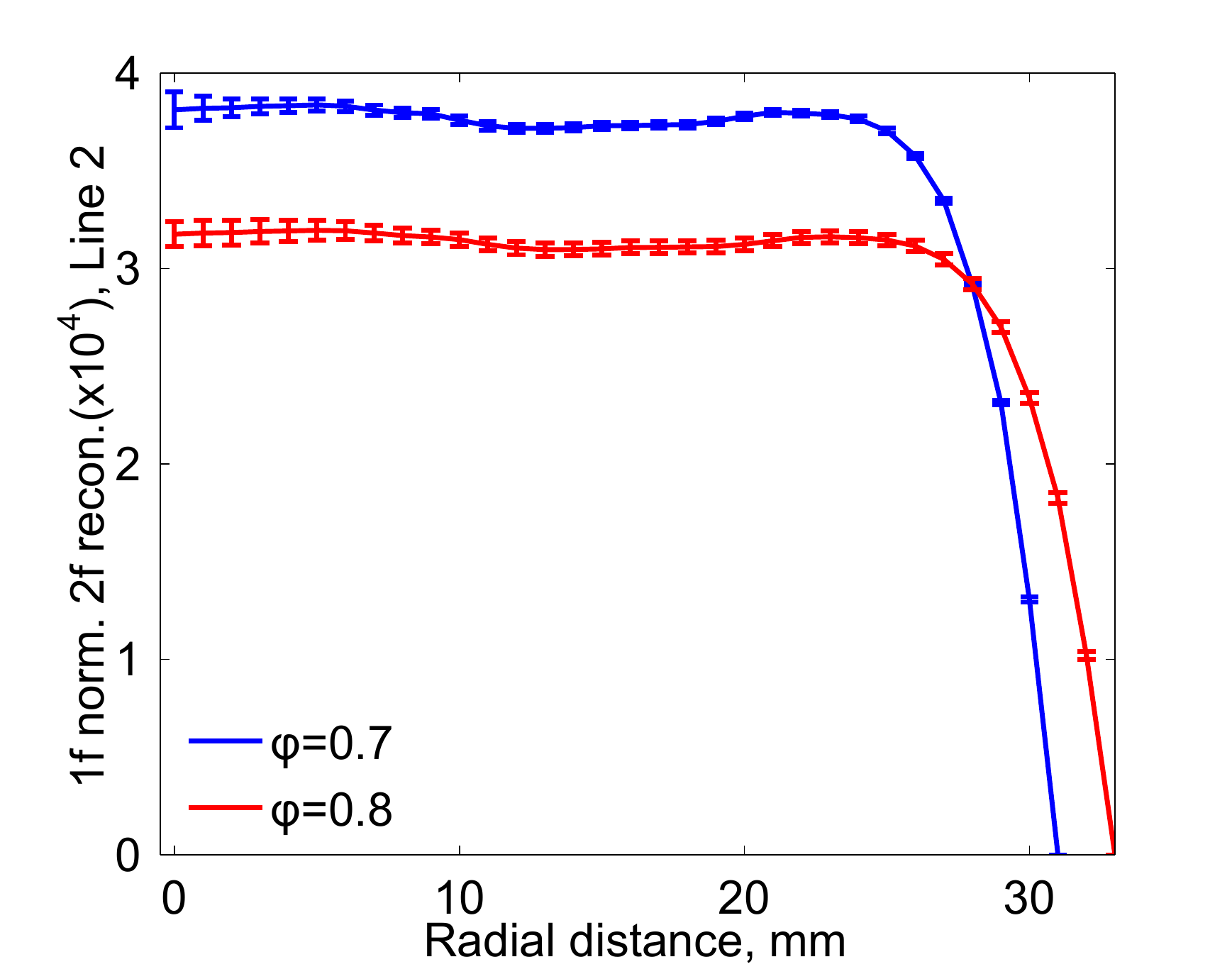}}\protect\caption{Reconstructed radial distributions of spectroscopic data at $\phi=0.7$
(blue) and $0.8$ (red). \label{fig:Result1D}}
\end{figure}
 
\begin{figure}
\begin{centering}
\subfloat[Reconstructed temperature profiles.\label{fig:temp}]{\centering{}\includegraphics[width=0.9\columnwidth]{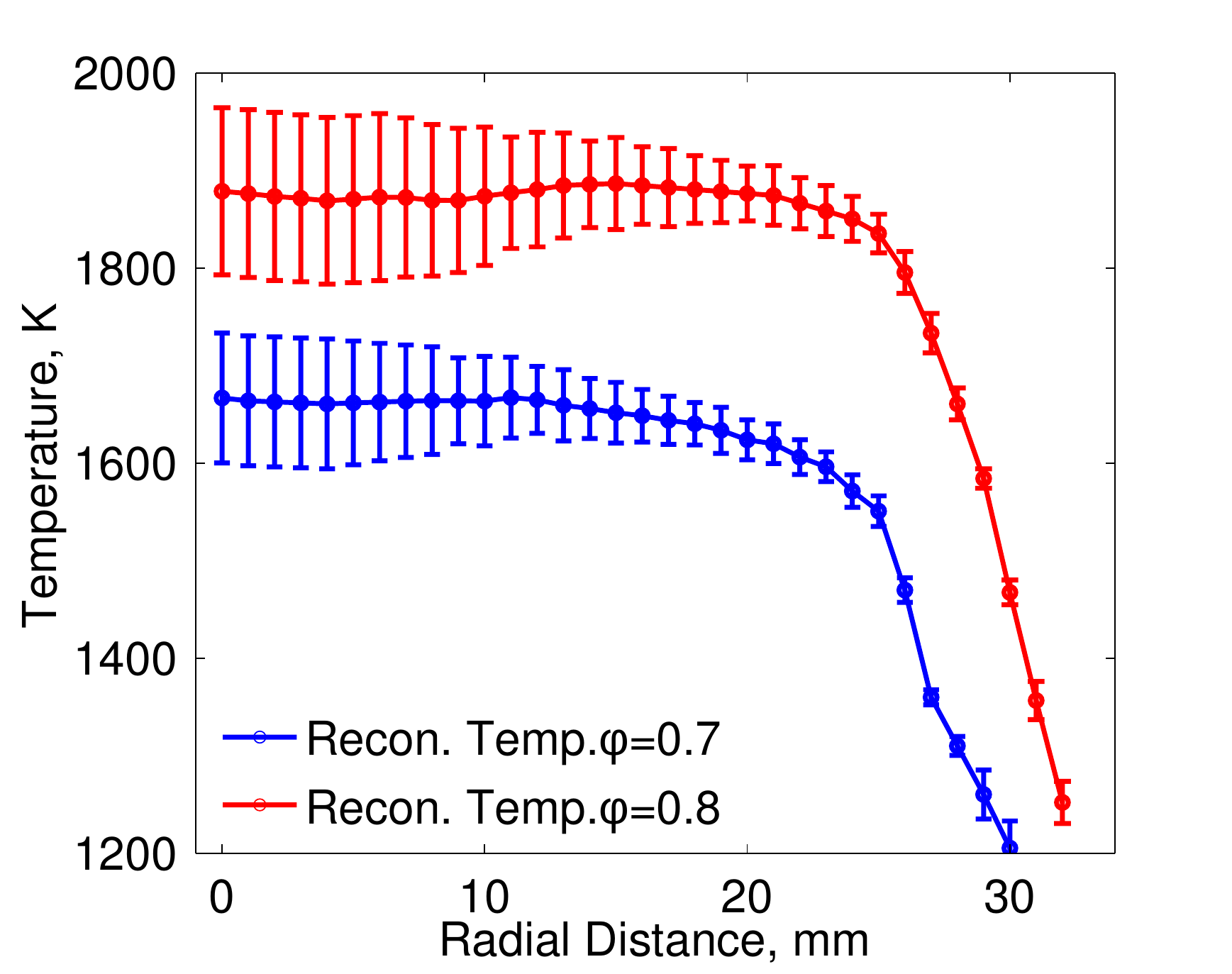}}
\par\end{centering}

\centering{}\subfloat[Reconstructed mole fraction profiles.\label{fig:conc}]{\centering{}\includegraphics[width=0.9\columnwidth]{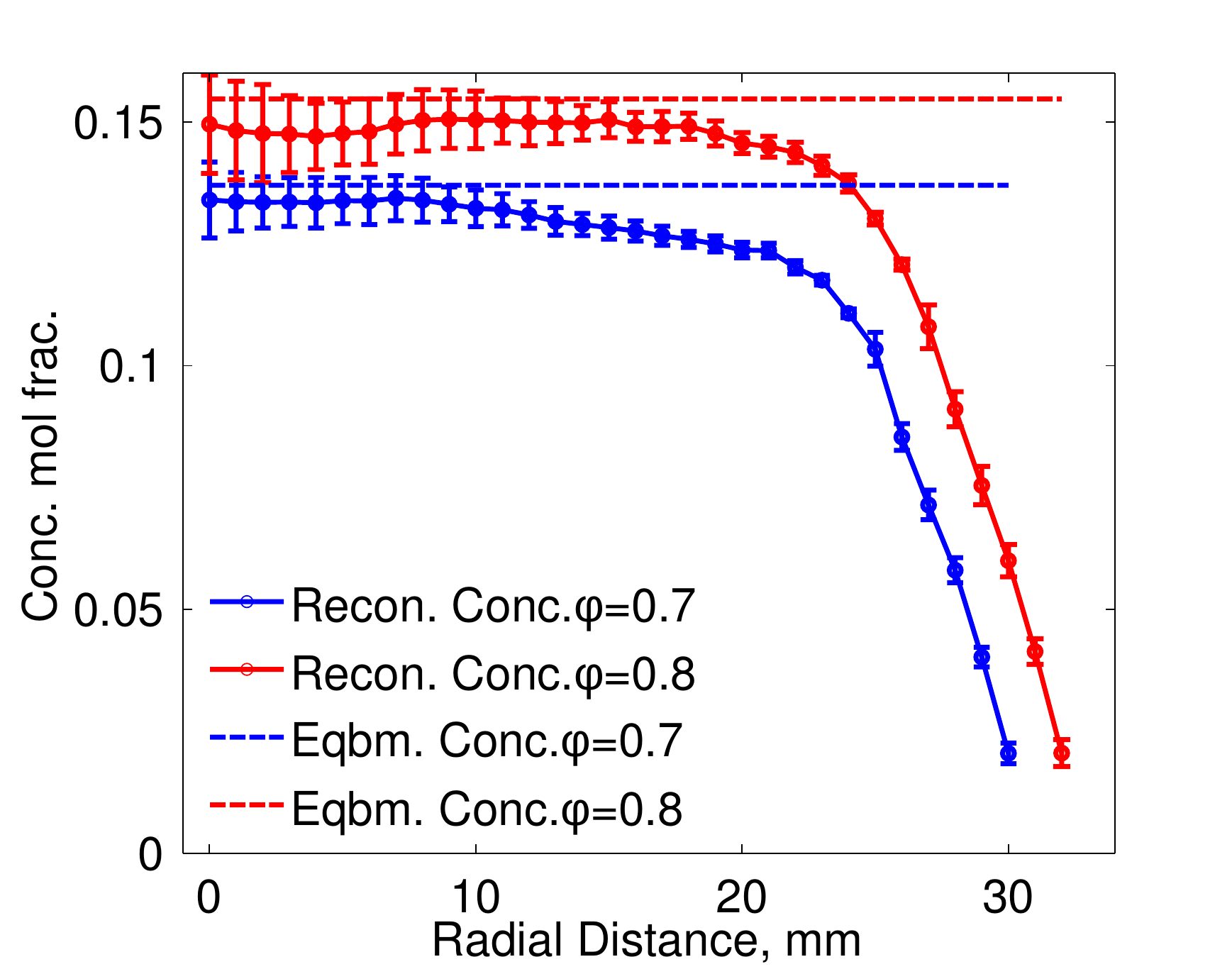}}\protect\caption{Temperature and mole fraction results with different values of $\phi$.
\label{fig:TCResults}}
\end{figure}
Tomographic reconstructions of radial temperature and mole fraction
profiles are obtained for two specific cases, where equivalence ratios
$\phi=0.7$ and $0.8$, and a linear velocity of 20 cm/sec are chosen.

\paragraph*{Individual Line Reconstructions.}

For tomographic measurements, spectroscopic data for both transitions
are recorded at equispaced radial locations and passed through a digital
lock-in amplifier with appropriately chosen detection phase $\psi_{d}$
(Section \ref{sub:Signal-Processing}). The $S_{\mathit{2f}}/S_{\mathit{1f}}$
signal is formed (Eq. \ref{eq:1fnorm2f}) and corrected for RAM (Eq.
\ref{eq:RAM}) and absorption outside the flame zone. The resulting
data represent line-of-sight projection values, and are shown in Figures
\ref{fig:proj_line1} and \ref{fig:proj_line2}. It is noted that
small uncertainty levels of measured projection values are due to
noise rejection that is intrinsic to WMS-2f. Spectroscopic projection
data are subsequently deconvolved using Tikhonov regularized Abel
inversion (Eq. \ref{eq:tikhonov}). The resulting distributions form
$\zeta_{j}(T,X)$, which are \emph{1f} normalized \emph{2f} signal
contributions per unit length within an axisymmetric domain.

Figures \ref{fig:line1} and \ref{fig:line2} show reconstructions
of $\zeta_{i}$ distributions due to $\ce{H_{2}O}$ absorption at
1442.67 and 1391.67 nm, respectively. Results clearly show the expected
behavior of a flat flame, i.e. a radial \emph{top-hat} distribution.
The central constant region is retrieved with good accuracy, whereas
the lateral wings with rapidly decreasing values result from the interaction
of the core region and the shroud flow. In measurements, non-zero
absorption is observed outside the core zone, which is attributed
to mixing as well as flow interactions due to thermal expansion of
reacting gases. This behavior was more pronounced for the higher equivalence
ratio; accordingly, radial measurement domains are extended to 31
and 33 mm for $\phi=0.7$ and $0.8$, respectively. No flame interaction
was detected at the outermost measurement point for either transition,
i.e. the radial locations lie beyond the edge of the flame.

\paragraph*{Temperature and Mole Fraction.}

Using individual line reconstructions, local temperature and water
vapor mole fractions are reconstructed point-by-point using a Levenberg-Marquardt
scheme (Eq. \ref{eq:LM}). Figures \ref{fig:temp} and \ref{fig:conc}
show resulting radial profiles for temperature and mole fraction,
respectively. Again, the expected top-hat profile is retrieved to
good accuracy. Near the center, temperatures and vapor mole fraction
at $\phi=0.7$ are reconstructed as 1670K and 0.135 mol/mol, respectively.
Corresponding values at $\phi=0.8$ are 1900 K and 0.15 mol/mol. Error
bars are based on measurement uncertainty that is propagated through
the tomographic reconstruction. Uncertainties near the center are
around $\pm70\mbox{\,\ensuremath{\mathrm{K}}}$ and $\mathrm{\pm}0.0077$
mol/mol at $\phi=0.7$, and almost $\pm90\,\mathrm{K}$ and $\pm0.01$
mol/mol at $\phi=0.8$. In Abel inversion, uncertainties are known
to grow towards the center due to the intrinsic ill-posedness of the
problem \citep{Daun2006}. This is clearly reflected in results; uncertainty
values are significantly smaller close to the edge of the flame.

\paragraph*{Validation. }

Mole fractions of water vapor near the center of the flame are expected
to be close to equilibrium mole fractions, regardless of parameters
such as velocity of reactants, matrix material of the burner and cooling
water flow rate and temperature \citep{Villarreal2005,DLR}. Results
in Figure \ref{fig:conc} illustrate that reconstructions near the
center are within 1.5\% and 4\% of equilibrium values for $\phi=0.7$
and $0.8$, respectively. The relative deterioration in mole fraction
results for the higher equivalence ratio is attributed to a decreased
sensitivity of the spectral line pair at higher temperatures.

In contrast to mole fraction values, reconstructed flame temperatures
in burner stabilized flames are not suited for validation purposes,
as they are always lower than adiabatic flame temperatures due to
heat transfer from the flame zone to the burner plate. Experimental
results are consistent, as adiabatic temperatures for $\phi=0.7$
and $0.8$ are $1839\,\mathrm{K}$ and $1997\,\mathrm{K}$, respectively.
Heat losses to the burner plate depend on a number of parameters,
e.g. flame stabilization height, flow velocity, burning velocity,
cooling water temperature and flow rates. Burner to burner temperature
differences of the order of 25 K have been documented for otherwise
identical experiments \citep{DLR}; furthermore, results can be affected
significantly by the burner material \citep{Migliorini2008}. While
a direct validation of temperature results is not justified, it is
noted that for a given temperature and mole fraction, spectral absorbance
and 2f heights are single valued within the temperature range of interest,
i.e. reconstructions yield unique solutions. Therefore, a validation
based on reconstructed mole fractions implies that temperatures are
reconstructed within good accuracy.

\section{Conclusions}

This work focuses tomographic TDLAS based on a two-transition technique,
which is an area where few studies exist. Within the optically thin
limit, it is shown that WMS-\emph{2f} signals obtained from line-of-sight
measurements represent path-integrated values. Thus, the 1f normalized
2f output of a lock-in-amplifier can be deconvolved using traditional
tomography; a variation of the three-point Abel algorithm is used
to reconstruct profiles of absorption data for two $\ce{H_{2}O}$
transitions. Averaging of multiple scans in combination with Tikhonov
regularization stabilizes the solution against detrimental effects
of measurement noise, which is exacerbated by the proximity of the
rays in traditional Abel inversion. Using reconstructions for two
lines, a Levenberg-Marquardt optimization routine is adopted to solve
simultaneously for unknown temperatures and mole fractions. This approach
represents an extension of the WMS-\emph{2f} technique and can be
easily extended to larger numbers of transitions. It is noted that
traditional WMS-\emph{2f} approaches fail in the context of tomography,
as line-shape integrals cannot be assumed to be constant. The experimental
technique was validated in experiments, where reconstructions of temperature
and mole fractions are in agreement with the expected behavior.

\section*{Acknowledgments}

This material is based upon work supported by the Louisiana Board
of Regents Research Competitiveness Subprogram under contract number
LEQSF(2010-2013)-RD-A-04. The authors thank Dr. Kyle Daun for valuable
discussions and insights.


\begin{thebibliography}{10}
\newcommand{\enquote}[1]{``#1''}

\bibitem{Philippe1993}
L.~Philippe and R.~Hanson, \enquote{Laser diode wavelength-modulation
  spectroscopy for simultaneous measurement of temperature, pressure, and
  velocity in shock-heated oxygen flows,} Appl. Opt. \textbf{32}, 6090--6103
  (1993).

\bibitem{Zhou2003}
X.~Zhou, X.~Liu, J.~B. Jeffries, and R.~K. Hanson, \enquote{Development of a
  sensor for temperature and water concentration in combustion gases using a
  single tunable diode laser,} Measurement Science and Technology \textbf{14},
  1459 (2003).

\bibitem{Zhou2005}
X.~Zhou, J.~Jeffries, and R.~Hanson, \enquote{Development of a fast temperature
  sensor for combustion gases using a single tunable diode laser,} Appl. Phys.
  B \textbf{81}, 711--722 (2005).

\bibitem{Farooq2008a}
A.~Farooq, J.~Jeffries, and R.~Hanson, \enquote{$\mathrm{CO_2}$ concentration
  and temperature sensor for combustion gases using diode-laser absorption near
  2.7 $\mu$m,} Appl. Phys. B \textbf{90}, 619--628 (2008).

\bibitem{silver1995}
J.~Silver, D.~Kane, and P.~Greenberg, \enquote{Quantitative species
  measurements in microgravity flames with near-ir diode lasers,} Appl. Opt.
  \textbf{34}, 2787--2801 (1995).

\bibitem{Villarreal2005}
R.~Villarreal and P.~Varghese, \enquote{Frequency-resolved absorption
  tomography with tunable diode lasers,} Appl. Opt. \textbf{44}, 6786--6795
  (2005).

\bibitem{Abel1826}
N.~Abel, \enquote{Aufl{\"o}sung einer mechanischen {A}ufgabe.} J. Reine Angew.
  Math. \textbf{1826}, 153--157 (1826).

\bibitem{Dasch1992}
C.~J. Dasch, \enquote{One-dimensional tomography: a comparison of abel,
  onion-peeling, and filtered backprojection methods,} Appl. Opt. \textbf{31},
  1146--1152 (1992).

\bibitem{Daun2006}
K.~Daun, K.~Thomson, F.~Liu, and G.~Smallwood, \enquote{Deconvolution of
  axisymmetric flame properties using {Tikhonov} regularization,} Appl. Opt.
  \textbf{45}, 4638--4646 (2006).

\bibitem{ma2008}
L.~Ma and W.~Cai, \enquote{Determination of the optimal regularization
  parameters in hyperspectral tomography,} Appl. Opt. \textbf{47}, 4186--4192
  (2008).

\bibitem{ma2008b}
L.~Ma and W.~Cai, \enquote{Numerical investigation of hyperspectral tomography
  for simultaneous temperature and concentration imaging,} Appl. Opt.
  \textbf{47}, 3751--3759 (2008).

\bibitem{cai2008}
W.~Cai, D.~Ewing, and L.~Ma, \enquote{Application of simulated annealing for
  multispectral tomography,} Comput. Phys. Commun. \textbf{179}, 250--255
  (2008).

\bibitem{Ma2009}
L.~Ma, W.~Cai, A.~W. Caswell, T.~Kraetschmer, S.~T. Sanders, S.~Roy, and J.~R.
  Gord, \enquote{Tomographic imaging of temperature and chemical species based
  on hyperspectral absorption spectroscopy,} Opt. Express \textbf{17},
  8602--8613 (2009).

\bibitem{hagen2007}
C.~Hagen and S.~Sanders, \enquote{Toward hyperspectral sensing in practical
  devices: Measurements of fuel, $\mathrm{H_2O}$ and gas temperature in a metal
  homogeneous charge compression ignition engine,} J. Near Infrared Spectrosc.
  \textbf{15}, 217 (2007).

\bibitem{cai2014b}
W.~Cai and C.~F. Kaminski, \enquote{Multiplexed absorption tomography with
  calibration-free wavelength modulation spectroscopy,} Appl. Phys. Letters
  \textbf{104}, 154106 (2014).

\bibitem{silver1992}
J.~A. Silver, \enquote{Frequency-modulation spectroscopy for trace species
  detection: theory and comparison among experimental methods,} Appl. Opt.
  \textbf{31}, 707--717 (1992).

\bibitem{Liu2004b}
J.~Liu, J.~Jeffries, and R.~Hanson, \enquote{Wavelength modulation absorption
  spectroscopy with 2f detection using multiplexed diode lasers for rapid
  temperature measurements in gaseous flows,} Appl. Phys. B \textbf{78},
  503--511 (2004).

\bibitem{Liu2005}
J.~C. Liu, G.~Rieker, J.~Jeffries, M.~Gruber, C.~Carter, T.~Mathur, and
  R.~Hanson, \enquote{Near-infrared diode laser absorption diagnostic for
  temperature and water vapor in a scramjet combustor,} Appl. Opt. \textbf{44},
  6701--6711 (2005).

\bibitem{Li2006}
H.~Li, G.~B. Rieker, X.~Liu, J.~B. Jeffries, and R.~K. Hanson,
  \enquote{Extension of wavelength-modulation spectroscopy to large modulation
  depth for diode laser absorption measurements in high-pressure gases,} Appl.
  Opt. \textbf{45}, 1052--1061 (2006).

\bibitem{Farooq2009b}
A.~Farooq, J.~Jeffries, and R.~Hanson, \enquote{Sensitive detection of
  temperature behind reflected shock waves using wavelength modulation
  spectroscopy of {$CO_2$} near 2.7 $\mu$m,} Appl. Phys. B \textbf{96},
  161--173 (2009).

\bibitem{Kluczynski1999}
P.~Kluczynski and O.~Axner, \enquote{Theoretical description based on fourier
  analysis of wavelength-modulation spectrometry in terms of analytical and
  background signals,} Appl. Opt. \textbf{38}, 5803--5815 (1999).

\bibitem{Liu2006}
X.~Liu, J.~Jeffries, R.~Hanson, K.~Hinckley, and M.~Woodmansee,
  \enquote{Development of a tunable diode laser sensor for measurements of gas
  turbine exhaust temperature,} Appl. Phys. B \textbf{82}, 469--478 (2006).

\bibitem{Rieker2009}
G.~B. Rieker, J.~B. Jeffries, and R.~K. Hanson, \enquote{Calibration-free
  wavelength-modulation spectroscopy for measurements of gas temperature and
  concentration in harsh environments,} Appl. Opt. \textbf{48}, 5546--5560
  (2009).

\bibitem{Hansen1987}
P.~C. Hansen, \emph{Rank-deficient and discrete ill-posed problems: numerical
  aspects of linear inversion}, vol.~4 (Society for Industrial Mathematics,
  1987).

\bibitem{Reid1981}
J.~Reid and D.~Labrie, \enquote{Second-harmonic detection with tunable diode
  lasers: Comparison of experiment and theory,} Appl. Phys. B \textbf{26},
  203--210 (1981).

\bibitem{Kluczynski2001}
P.~Kluczynski, {\AA}.~M. Lindberg, and O.~Axner, \enquote{Background signals in
  wavelength-modulation spectrometry by use of frequency-doubled diode-laser
  light. ii. experiment,} Appl. Opt. \textbf{40}, 794--805 (2001).

\bibitem{rothman2009}
L.~S. Rothman, I.~E. Gordon, A.~Barbe, D.~C. Benner, P.~F. Bernath, M.~Birk,
  V.~Boudon, L.~R. Brown, A.~Campargue, J.-P. Champion \emph{et~al.},
  \enquote{The {HITRAN} 2008 molecular spectroscopic database,} J. Quant.
  Spectrosc. Radiat. Transfer \textbf{110}, 533--572 (2009).

\bibitem{rothman2010}
L.~Rothman, I.~Gordon, R.~Barber, H.~Dothe, R.~Gamache, A.~Goldman,
  V.~Perevalov, S.~Tashkun, and J.~Tennyson, \enquote{{HITEMP}, the
  high-temperature molecular spectroscopic database,} J. Quant. Spectrosc.
  Radiat. Transfer \textbf{111}, 2139--2150 (2010).

\bibitem{DLR}
P.~Weigand, R.~L{\"u}ckerath, and W.~Meier, \enquote{Documentation of flat
  premixed laminar $\mathrm{CH_4}$air standard flames: Temperatures and species
  concentrations,} Tech. rep., DLR Institute (2003).

\bibitem{Migliorini2008}
F.~Migliorini, S.~De~Iuliis, F.~Cignoli, and G.~Zizak, \enquote{How flat is the
  rich premixed flame produced by your {McKenna} burner?} Combust. Flame
  \textbf{153}, 384--393 (2008).

\end{thebibliography}
\end{document}